\documentclass[9pt,twocolumn,twoside]{osajnl}
\usepackage{stfloats}
\usepackage{float}
\usepackage{subfigure}
\usepackage{graphicx}
\usepackage{amsmath}
\usepackage{amssymb}
\usepackage{multirow}
\usepackage{setspace}
\usepackage{mathrsfs}

\newcommand{\bb}[1]{{\mathbf{#1}}}
\newcommand{\nn}{\nonumber}
\usepackage{color}

\journal{josaa} 

\setboolean{shortarticle}{false}

\ifthenelse{\boolean{shortarticle}}{\colorlet{color2}{color2b}}{\colorlet{color2}{color2}} 

\title{Fast and Accurate Computation of Normalized Bargmann Transform}

\author[1,*]{Soo-Chang Pei}
\author[2]{Shih-Gu Huang}

\affil[1]{Department of Electrical Engineering \& Graduate Institute of Communication Engineering, National Taiwan University, Taipei 10617, Taiwan}
\affil[2]{Graduate Institute of Communication Engineering, National Taiwan University, Taipei 10617, Taiwan}

\affil[*]{Corresponding author: peisc@ntu.edu.tw}

\dates{Compiled December 29, 2016}

\ociscodes{(070.2590) ABCD transforms; (070.2580) Paraxial wave optics; (270.0270) Quantum optics; (070.0070) Fourier optics and signal processing; (070.2575) Fractional Fourier transforms.}

\doi{\url{http://dx.doi.org/10.1364/ao.XX.XXXXXX}}

\begin{abstract}
Linear canonical transform (LCT) was extended to complex-valued parameters, called complex LCT,
to describe the complex amplitude propagation through lossy or lossless optical systems.
Bargmann transform is a special case of the complex LCT.
In this paper, we normalize the Bargmann transform such that it can be bounded near infinity.
We derive the relationships of the normalized Bargmann transform to  Gabor transform,  Hermite Gaussian functions, gyrator transform and 2D nonseparable LCT.
Several kinds of fast and accurate computational methods of the normalized Bargmann transform and its inverse are proposed based on these relationships.
\normalfont {\small{© 2016 Optical Society of America. One print or electronic copy may be made for personal use only. Systematic reproduction and distribution, duplication of any material in this paper for a fee or for commercial purposes, or modifications of the content of this paper are prohibited.}}
\end{abstract}

\setboolean{displaycopyright}{false}

\begin{document}

\maketitle
\thispagestyle{fancy}

\ifthenelse{\boolean{shortarticle}}{\ifthenelse{\boolean{singlecolumn}}{\abscontentformatted}{\abscontent}}{}

\allowdisplaybreaks   
\section{Introduction}\label{sec:Intro}
Linear canonical transform (LCT), first introduced in \cite{collins1970lens,moshinsky1971linear}, is an important tool to describe paraxial light propagation through first-order optical systems \cite{wolf1979integral,nazarathy1982first,bastiaans1989propagation,ozaktas2001fractional}.
It is also useful and attractive in many signal processing applications
\cite{barshan1997optimal,pei2001relations,bastiaans2003phase,hennelly2005optical,sharma2006signal,pei2013reversible,pei2015fast}.
Some famous transforms such as Fourier transform,  fractional Fourier transform (FRFT) and Fresnel transform are the special cases of the LCT \cite{ozaktas2001fractional,ding2001research,pei2002eigenfunctions}.

The LCT was extended to complex-valued parameters in \cite{wolf1974canonicalI,wolf1974canonicalII,kocc2010fastCLCT,liu2011digital,wolf2013integral}, called \emph{complex LCT} for simplification whereas the LCT with real parameters is called \emph{real LCT}.
The real LCT is a unitary mapping of the Hilbert space 
$\mathscr{L}^2(\mathscr{R})$  onto itself, while the complex LCT is a unitary transformation between $\mathscr{L}^2(\mathscr{R})$ and the Bargmann-like space ${\mathscr{B}}_M$.
The definition of the complex LCT is the same as that of the real LCT:
\begin{align}\label{eq:Intro04}
{\cal O}_{\rm CLCT}^{\bb M}\{s(t)\}
= \left\{
  \begin{array}{l l}
\sqrt {\frac{1}{{j2\pi b}}} \int\limits {{e^{j\left(\frac{d}{{2b}}{u^2} - \frac{1}{b}ut + \frac{a}{{2b}}{t^2}\right)}}} s(t)\;dt,&  b \ne 0\\
\sqrt d {e^{j\frac{cd}{2}{u^2}}}s(du),& b = 0
  \end{array} \right.,
\end{align}
except that the parameters are complex-valued:
\begin{align}\label{eq:Intro08}
\bb M = \begin{bmatrix}
a\ &b\\
c\ &d
\end{bmatrix}= \begin{bmatrix}
{{a_r} + j{a_i}}&{{b_r} + j{b_i}}\\
{{c_r} + j{c_i}}&{{d_r} + j{d_i}}
\end{bmatrix}\quad {\rm and} \quad ad - bc = 1.
\end{align}
When $\bb M=(0,\ j;\ j,\ 0)$, the complex LCT reduces to the well-known bilateral Laplace transform \cite{wolf1977self,torre2003linear,davies2012integral}.
Other special cases include Bargmann transform, Gauss-Weierstrass transform \cite{bargmann1961hilbert,wolf1974canonicalI,wolf1977self,wolf2013integral}, complex-ordered FRFT \cite{shih1995optical,bernardo1996optical,bernardo1997talbot} and fractional Laplace transform \cite{torre2003linear}.
The complex LCT has been used to describe the complex amplitude propagation through several kinds of lossy or lossless optical systems \cite{nazarathy1982first,bastiaans2006first,kocc2010fastCLCT}.



In this paper, we focus on the Bargmann transform,
which is introduced by V. Bargmann in 1961 \cite{bargmann1961hilbert} to map functions
from $\mathscr{L}^2(\mathscr{R})$  onto the Bargmann space ${\mathscr{B}}_B$ (also known as Bargmann–Segal-Foch space).
The Bargmann transform has been applied to quantum field theory \cite{bargmann1961hilbert}, quantum optics \cite{klauder2006fundamentals} as well as signal processing on phase space \cite{folland2016harmonic}.
However, the output of the Bargmann transform is unbounded near infinity.
Thus, in this paper, we normalize the Bargmann transform.

Since the Bargmann transform is a special case of the complex LCT,
one can use the existing algorithms of the complex LCT to compute the Bargmann transform, and then normalize it to obtain the normalized Bargmann transform.
Digital computation of the real LCT has been widely discussed in
\cite{hennelly2005generalizing,hennelly2005fast,ozaktas2006efficient,koc2008digital,pei2011discrete,pei2015fast,kocc2016fast}.
Ko{\c{c}} \emph{et al}. extended the computation of the real LCT to the complex LCT \cite{kocc2010fastCLCT}.
Liu \emph{et al}. discretized the complex LCT directly by sampling  \cite{liu2011digital}.
However, the Bargmann transform may have very large energy in the boundary, causing computation difficulty and inaccuracy in real world.
Accordingly, we want to develop computational methods to obtain the normalized Bargmann transform directly.

The relationships of the Bargmann transform to Gabor transform and Hermite Gaussian (HG) functions have been discussed in \cite{hall1994segal,fan2002epr,abreu2010sampling,zayed2013chromatic}.
Based on these relationships, we develop two computational methods of the normalized Bargmann transform.
One is based on the Gabor transform, while another one is based on HG expansion and synthesis.
Furthermore, we derive its connections with
gyrator transform \cite{rodrigo2007gyrator,pei2009properties,pei2015discrete} and 2D nonseparable LCT \cite{folland1989harmonic,pei2001two,alieva2005alternative,pei2016two}, followed by two more computational methods of the  normalized Bargmann transform.
We also derive several computational methods for the inverse normalized Bargmann transform.


\section{Definition of Normalized Bargmann Transform}\label{sec:Mod}
The Bargmann transform is a special case of the complex LCT.
When the parameter matrix in (\ref{eq:Intro08}) is given by
\begin{align}\label{eq:Intro24}
\bb M = \begin{bmatrix}
\frac{1}{\sqrt 2} &-j\frac{1}{\sqrt 2}\\
-j\frac{1}{\sqrt 2}&\frac{1}{\sqrt 2}
\end{bmatrix},
\end{align}
the complex LCT in (\ref{eq:Intro04}) becomes the Bargmann transform, denoted by ${\cal B}$,
\begin{align}\label{eq:Intro20}
{S_{\rm B}}(z) ={\cal B}\!\left\{s(t)\right\} =  {2^{ - \frac{1}{4}}}{\pi ^{ - \frac{1}{2}}}\int\limits_{ - \infty }^\infty  {{e^{ - \frac{{{z^2}}}{2}+ \sqrt 2 zt - \frac{{{t^2}}}{2} }}s(t)} dt,
\end{align}
where $z$ is in the complex plane $\mathbb C$.
The weighting function of the Bargmann transform is \cite{fan2002epr,wolf2013integral}
\begin{align}\label{eq:Intro32}
w(z) = {2^{\frac{1}{2}}}{\pi ^{ - \frac{1}{2}}}{e^{ - |z{|^2}}}.
\end{align}
Accordingly, the inverse Bargmann transform, denoted by ${\cal B}^{-1}$, is given by
\begin{align}\label{eq:Intro36}
s(t)={\cal B}^{-1}\!\left\{{S_{\rm B}}(z)\right\} & =  {2^{ - \frac{1}{4}}}{\pi ^{ - \frac{1}{2}}}\!\!\int_{\mathbb{C}}w(z) {{{\left( {{e^{ - \frac{{{z^2}}}{2} + \sqrt 2 zt - \frac{{{t^2}}}{2}}}} \right)}^{\!\!*}}}\! {S_{\rm B}}(z)dz\nn\\
&= {2^{\frac{1}{4}}}{\pi ^{ - 1}}\!\!\int_{\mathbb{C}} {{e^{- |z|^2 - \frac{{{ {\overline z}^2}}}{2} + \sqrt 2 {\overline z}t - \frac{{{t^2}}}{2} }}} {S_{\rm B}}(z)dz,
\end{align}
where $\overline z$ is the complex conjugate of $z$.
The Bargmann transform maps functions from the real line to the complex plane.
Thus, assume $z=x+jy$, and then
we can express the Bargmann transform  as a 2D transform with real arguments $x$ and $y$:
\begin{align}\label{eq:Intro40}
{S_{\rm B}}(x,y)&={\cal B}\!\left\{s(t)\right\} \nn\\
& =  {2^{ - \frac{1}{4}}}{\pi ^{ - \frac{1}{2}}}\int\limits_{ - \infty }^\infty  {{e^{ - \frac{{{{(x + jy)}^2}}}{2} + \sqrt 2 (x + jy)t - \frac{{{t^2}}}{2}}}s(t)} dt,
\end{align}
and similarly the inverse transform in (\ref{eq:Intro36}) becomes
\begin{align}\label{eq:Intro44}
s(t)&={\cal B}^{-1}\!\left\{{S_{\rm B}}(x,y)\right\} \nn\\
& =  {2^{\frac{1}{4}}}{\pi ^{ - 1}}\int\limits_{ - \infty }^\infty  {\int\limits_{ - \infty }^\infty  {{e^{ - {x^2} - {y^2} - \frac{{{{(x - jy)}^2}}}{2} + \sqrt 2 (x - jy)t - \frac{{{t^2}}}{2}}}} } {S_{\rm B}}(x,y)dxdy.
\end{align}
However, the Bargmann transform  may be 
unbounded when $x$ or $y$ approaches  infinity. 
Accordingly, we normalize the Bargmann transform by
portioning out the weighting in (\ref{eq:Intro32}) equally to the forward transform and the inverse transform.
The normalized Bargmann transform, denoted by ${\cal N\!B}$, is defined as
\begin{align}\label{eq:Mod02}
S_{\rm NB}(z)&={\cal N\!B}\!\left\{s(t)\right\}  =  \sqrt {w(z)} {S_{\rm B}}(z)\nn\\
&= {2^{\frac{1}{4}}}{\pi ^{ - \frac{1}{4}}}{e^{ - {{\frac{{|z|^2}}{2}}}}}{S_{\rm B}}(z)\\
&= {\pi ^{ - \frac{3}{4}}}\int\limits_{ - \infty }^\infty  {{e^{ - {{\frac{{|z|^2}}{2}}} - \frac{{{z^2}}}{2} + \sqrt 2 zt - \frac{{{t^2}}}{2}}}s(t)} dt.\label{eq:Mod04}
\end{align}
Compared with the complex LCT defined in (\ref{eq:Intro04}), we can find out that the normalized Bargmann transform is not a special case of the complex LCT because of the term $|z|^2=zz^*$.
Substituting the relationship in (\ref{eq:Mod02}) to the inverse Bargmann transform in (\ref{eq:Intro36}), we have the inverse normalized Bargmann transform ${\cal N\!B}^{-1}$ given by
\begin{align}\label{eq:Mod08}
s(t)={\cal N\!B}^{-1}\!\left\{{S_{\rm NB}}(z)\right\}
= {\pi ^{ - \frac{3}{4}}}\int_{\mathbb{C}} {{e^{- {{\frac{{|z|}}{2}}^2} - \frac{{{{\overline z}^2}}}{2} + \sqrt 2 \overline zt - \frac{{{t^2}}}{2}}}} S_{\rm NB}(z)dz.
\end{align}
The normalized Bargmann transform can also be expressed as a 2D transform; that is, (\ref{eq:Mod04}) can be rewritten as
\begin{align}
S_{\rm NB}(x,y)&={\cal N\!B}\!\left\{s(t)\right\}\nn\\ 
&= {\pi ^{ - \frac{3}{4}}}\int\limits_{ - \infty }^\infty  {{e^{ - {x^2} - jxy + \sqrt 2 (x + jy)t - \frac{{{t^2}}}{2}}}s(t)} dt,\label{eq:Mod16}
\end{align}
and the inverse normalized Bargmann transform in (\ref{eq:Mod08}) can be rewritten as
\begin{align}
s(t)&={\cal N\!B}^{-1}\!\left\{{S_{\rm NB}}(x,y)\right\}\nn\\
&= {\pi ^{ - \frac{3}{4}}}\int\limits_{ - \infty }^\infty  {\int\limits_{ - \infty }^\infty  {{e^{ - {x^2} + jxy + \sqrt 2 (x - jy)t - \frac{{{t^2}}}{2}}}} } S_{\rm NB}(x,y)dxdy.\label{eq:Mod20}
\end{align}
Comparing (\ref{eq:Mod16}) and (\ref{eq:Mod20}), we can find out that the kernel in the inverse transform is just the complex conjugate of the kernel in the forward transform.
Thus, the normalized Bargmann transform is a unitary transform.

Besides implementing the normalized Bargmann transform and its inverse directly by (\ref{eq:Mod16}) and (\ref{eq:Mod20}), some other computational methods are proposed in the following sections.
We don't pay much attention to the conventional (i.e. unnormalized) Bargmann transform because it can be easily calculated from
\begin{align}\label{eq:Pro04}
{S_{\rm B}}(x,y)={2^{-\frac{1}{4}}}{\pi ^{  \frac{1}{4}}}{e^{  \frac{{{x^2} + {y^2}}}{2}}}S_{\rm NB}(x,y),
\end{align}
once the normalized Bargmann transform is obtained.


\section{Computation Based on Gabor Transform}\label{sec:Gabor}
The Bargmann transform is closely connected to the Gabor transform \cite{hall1994segal,abreu2010sampling}.
In this section, we derive the computations of the normalized Bargmann transform and its inverse based on this relationship.

The Gabor transform is one of the most popular short-time Fourier transforms that uses Gaussian function with unit variance as the window function:
\begin{align}\label{eq:Gabor08}
{G
}(\tau ,\omega ) = \frac{1}{{\sqrt {2\pi } }}\int\limits_{ - \infty }^\infty  {s(t){e^{ - \frac{1}{2}{{(\tau  - t)}^2}}}} {e^{ - j\omega t}}dt,
\end{align}
and its connection with the Bargmann transform is given by
\begin{align}\label{eq:Gabor10}
G\!\left( {\tau, \omega} \right)=
{2^{-\frac{1}{4}}}e^{-j\frac{\tau\omega}{2}}e^{-\frac{{{\tau^2} + {\omega^2}}}{4}}{S_{\rm B}}\left( {\frac{\tau}{\sqrt 2}, -\frac{\omega}{\sqrt 2}} \right).
\end{align}
That is,
\begin{align}\label{eq:Gabor20}
{S_{\rm B}}(x,y) = {2^{\frac{1}{4}}}{e^{ - jxy}}{e^{\frac{{{x^2} + {y^2}}}{2}}}G\!\left( {\sqrt 2 x, - \sqrt 2 y} \right).
\end{align}
Because of the term ${e^{\frac{{{x^2} + {y^2}}}{2}}}$, it is obvious that  the Bargmann transform is unbounded if the Gabor transform isn't close to zero when $x$ or $y$ approaches  infinity.

\begin{figure}[t]
\centering
\includegraphics[width=0.9\columnwidth,clip=true]{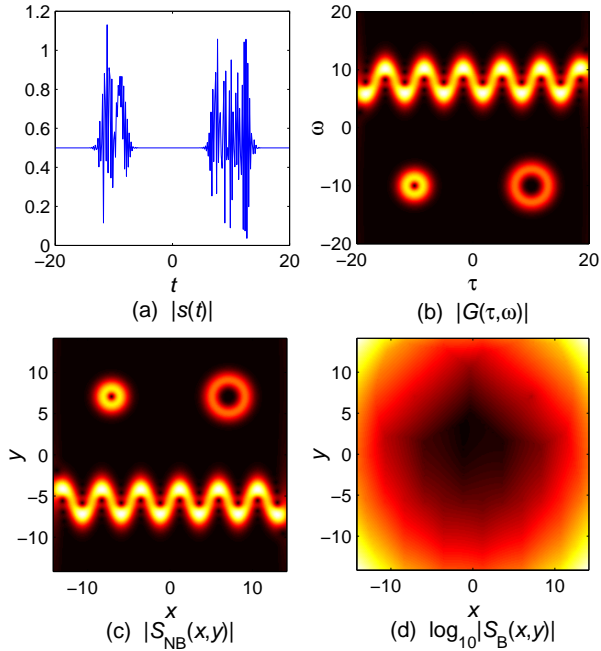}
\vspace*{-4pt}
\caption{
Digital Computation based on Gabor transform: the envelopes of (a) input signal $s(t)$, (b) Gabor transform ${G}(\tau ,\omega )$, (c) normalized Bargmann transform ${S_{\rm NB}}(x,y)$ and (d) 
unnormalized Bargmann transform ${S_{\rm B}}(x,y)$, where
the sampling periods are $\Delta_\tau=\Delta_\omega=\Delta_t=0.157$ and $\Delta_x=\Delta_y=\frac{0.157}{\sqrt 2}$.
The unnormalized Bargmann transform is plotted in logarithmic scale because of the very large energy in the boundary.
}
\label{fig:Gabor}
\end{figure}

Recall the relation in (\ref{eq:Pro04}).
The normalized Bargmann transform is just the Gabor transform multiplied by some phase term, i.e.
\begin{align}\label{eq:Gabor16}
S_{\rm NB}(x,y) = {2^{\frac{1}{2}}}{\pi ^{ - \frac{1}{4}}}{e^{ - jxy}}\,G\!\left( {\sqrt 2 x, - \sqrt 2 y} \right).
\end{align}
Therefore, if the input signal has bounded time-frequency energy distribution, the output of the normalized Bargmann transform is also bounded.
For digital computation, the Gabor transform can be realized by several different approaches such as the FFT-based algorithm and the chirp-Z-based algorithm.
But if the normalized Bargmann transform requires output sampling periods being $\Delta_x$ and $\Delta_y$ for $x$ and $y$, respectively, the sampling periods of the Gabor transform should be $\Delta_\tau={\sqrt 2}\Delta_x$ and $\Delta_\omega={\sqrt 2}\Delta_y$.
An example is given in Fig.~\ref{fig:Gabor} where the input
signal $s(t)$
consists of one sinusoidal FM signal and two Hermite Gaussian (HG) functions, which will be defined in the next section.
The envelope of $s(t)$
with sampling period $\Delta_t=0.157$
is shown in Fig.~\ref{fig:Gabor}(a).
The envelope of the Gabor transform ${G}(\tau ,\omega )$
computed by FFT with sampling periods $\Delta_\tau=\Delta_\omega=0.157$
is shown in Fig.~\ref{fig:Gabor}(b).
With $\Delta_x=\Delta_y=\frac{0.157}{\sqrt 2}$,
the normalized Bargmann transform ${S_{\rm NB}}(x,y)$ and the unnormalized Bargmann transform ${S_{\rm B}}(x,y)$ calculated from (\ref{eq:Gabor20}) and (\ref{eq:Gabor16}) have envelopes depicted in Figs.~\ref{fig:Gabor}(c) and (d), respectively.
The unnormalized Bargmann transform has very large energy in the boundary 
and thus is plotted in logarithmic scale.

The Gabor transform has the following recovery property:
\begin{align}\label{eq:Gabor28}
s(t) = \frac{1}{{2\pi }}\int\limits_{ - \infty }^\infty  {\int\limits_{ - \infty }^\infty  {{G }(\tau ,\omega )} {e^{ j\omega t}}d\tau }\, d\omega .
\end{align}
With the relationship in (\ref{eq:Gabor16}), the above equation leads to the inverse normalized Bargmann transform given by
\begin{align}\label{eq:Gabor32}
s(t) &= \frac{1}{\pi }\int\limits_{ - \infty }^\infty  {\int\limits_{ - \infty }^\infty  {{G}\left( {\sqrt 2 x, - \sqrt 2 y} \right)}{e^{-j\sqrt 2 yt}}dx}\, dy\nn\\
 &= {2^{ - \frac{1}{2}}}{\pi ^{ - \frac{3}{4}}}\int\limits_{ - \infty }^\infty \left( \int\limits_{ - \infty }^\infty  {S_{\rm NB}(x,y)}  {e^{jxy}}dx\right) {e^{-j\sqrt 2 yt}} dy.
\end{align}
We can find out that the above inverse transform is simpler than that in (\ref{eq:Mod20}).
For digital computation, we only require a 2D pointwise product for $e^{jxy}$, a 2D summation along $x$ axis , and a 1D FFT along $y$ axis.
The Gabor transform has another recovery property; that is,
\begin{align}\label{eq:Gabor36}
 s(\tau)= \frac{1}{\sqrt {2\pi }}\int\limits_{ - \infty }^\infty  {{G }(\tau,\omega )} {e^{j\omega \tau}}d\omega.
\end{align}
From (\ref{eq:Gabor16}) and above equation with $\tau=\sqrt{2}x$ and $\omega=-\sqrt{2}y$,
the input signal can be recovered by a different way:
\begin{align}\label{eq:Gabor40}
s\left( {\sqrt 2 x} \right) &= \sqrt {\frac{1}{\pi }} \int\limits_{ - \infty }^\infty  {{G}\left( {\sqrt 2 x, - \sqrt 2 y} \right)} {e^{ - j2xy}}dy\nn\\
 &= {2^{ - \frac{1}{2}}}{\pi ^{ - \frac{1}{4}}} \int\limits_{ - \infty }^\infty  {S_{\rm NB}(x,y)}\, {e^{ - jxy}}dy.
\end{align}
Compared with (\ref{eq:Gabor32}), the above inverse transform requires only one integral.
That is, the digital computation of this inverse transform doesn't require the 1D FFT used in (\ref{eq:Gabor32}).
However, the cost of lower complexity is that the sampling period of the recovered signal depends on the sampling period the normalized Bargmann transform, i.e. $\sqrt{2}\Delta_x$.
As an example, we recover the signal $s(t)$ in Fig.~\ref{fig:Gabor}(a) from ${S_{\rm NB}}(x,y)$ in Fig.~\ref{fig:Gabor}(c) using the two kinds of inverse transforms  (\ref{eq:Gabor32}) and (\ref{eq:Gabor40}).
Both methods have perfect recovery with normalized mean-square error (NMSE) smaller than $10^{-27}$.

Given the sampled input $s[n] \buildrel \Delta \over = s(n{\Delta _t})$ and the recovered input $s_r[n]$, the NMSE is defined as
\begin{align}\label{eq:Gabor46}
{\rm{NMSE}} = \frac{{\sum\limits_n^{} {{{\left| {s[n] - {s_r}[n]} \right|}^2}} }}{{\sum\limits_n^{} {{{\left| {s[n]} \right|}^2}} }},
\end{align}
while the MSE is defined as
\begin{align}\label{eq:Gabor48}
{\rm{MSE}} = \frac{1}{N}\sum\limits_n^{} {{{\left| {s[n] - {s_r}[n]} \right|}^2}}.
\end{align}
The NMSE is used in this paper because it won't be affected by the increase of signal energy.

\section{Computation Based on Hermite Gaussian Expansion and Synthesis}\label{sec:HG}
The Hermite Gaussian (HG) function of order $n$ is defined as
\begin{align}\label{eq:HG04}
HG_{n}(t) = {\left( {2^{n }}n!\sqrt{\pi} \right)^{-\frac{1}{2}}}{e^{ - \frac{t^2}{2}}}{H_n}(t),
\end{align}
where $H_n(t)$ is the physicists' Hermite polynomial.
It has been shown in \cite{fan2002epr,zayed2013chromatic} that the Bargmann transform of the $n$th-order HG function is given by
\begin{align}\label{eq:HG06}
{\cal B}\!\left\{H{G_n}(t)\right\}=\left( {2\pi} \right)^{ - \frac{1}{4}}\left( {n!} \right)^{ - \frac{1}{2}}{z^n}.
\end{align}
According to (\ref{eq:Mod02}) and letting $z=x+jy$,
the normalized Bargmann transform of the HG function is given by
\begin{align}\label{eq:HG32}
{\cal N\!B}\!\left\{H{G_n}(t)\right\}={\left( {\pi n!} \right)^{ - \frac{1}{2}}}{(x + jy)^n}{e^{ - \frac{{{x^2} + {y^2}}}{2}}}.
\end{align}
The Lagueree Gaussian (LG) function \cite{beijersbergen1993astigmatic} is defined as
\begin{align}\label{eq:HG36}
L{G_{m,n}}(\rho,\phi ) = C_{mn}\,{\rho^{|m - n|}}{e^{ - j(m - n)\phi }}L_{\min (m,n)}^{|m - n|}({\rho^2}){e^{ - \frac{{{\rho^2}}}{2}}},
\end{align}
where $L_p^l( \cdot )$ is the associated Lagueree polynomial, and the constant $C_{mn}$ is given by
\begin{align}\label{eq:HG38}
C_{mn}={( - 1)^{\min (m,n)}}\frac{{\min (m,n)!}}{{\sqrt {\pi m!n!} }}.
\end{align}
Assume $x+jy=\rho e^{j\phi}$ and let $m=0$.
Then, the LG function in (\ref{eq:HG36}) becomes
\begin{align}\label{eq:HG40}
L{G_{0,n}}(x,y)
&= {( - 1)^0}\frac{{0!}}{{\sqrt {\pi 0!n!} }}{\rho^n}{e^{jn\phi }}L_0^n({\rho^2}){e^{ - \frac{{{\rho^2}}}{2}}}\nn\\
& = \frac{1}{{\sqrt {\pi n!} }}{(x + jy)^n}{e^{ - \frac{{{x^2} + {y^2}}}{2}}}.
\end{align}
From (\ref{eq:HG32}) and (\ref{eq:HG40}), one has
\begin{align}\label{eq:HG44}
{\cal N\!B}\!\left\{H{G_n}(t)\right\}=L{G_{0,n}}(x,y).
\end{align}
The reason of expressing (\ref{eq:HG44}) by the LG function is that the LG function can be computed by the HG functions.
There are fast algorithms to generate discrete HG functions.
Then, the digital computation of the normalized Bargmann transform can be completed realized by the discrete HG functions.
Another reason is that  equation (\ref{eq:HG44}) leads to the connection between the normalized Bargmann transform and the gyrator transform, which will be discussed in the next section.

The HG functions can form an orthonormal basis, and thus any signal $s(t)$ can be expanded in terms of the HG functions, i.e.
\begin{align}\label{eq:HG52}
s(t) = \sum\limits_{n = 0}^\infty  {{{\widehat s}_n}} H{G_n}(t),
\end{align}
where the expansion coefficients ${{\widehat s}_n}$'s are given by
\begin{align}\label{eq:HG54}
{{\widehat s}_n} = \int\limits_{ - \infty }^\infty  {s(t)} H{G_n}(t)dt.
\end{align}
According to (\ref{eq:HG44}), performing the normalized Bargmann transform to the both sides of (\ref{eq:HG52}) leads to
\begin{align}\label{eq:HG56}
S_{\rm NB}(x,y)
= \sum\limits_{n = 0}^\infty  {{{\widehat s}_n}}{\cal N\!B}\!\left\{H{G_n}(t)\right\}=\sum\limits_{n = 0}^\infty  {{{\widehat s}_n}}L{G_{0,n}}(x,y).
\end{align}
That is, expanding the input signal by the HG functions with coefficients ${{{\widehat s}_n}}$'s,
the output of the normalized Bargmann transform can be synthesized by the LG functions with the same coefficients ${{{\widehat s}_n}}$'s.
Equation (\ref{eq:HG56}) also implies that all the outputs of the normalized Bargmann transform are in the space formed by the LG functions $LG_{0,n}$'s.
Next, consider the inverse transform.
Because the LG functions are orthonormal to each other, according to (\ref{eq:HG56}), the coefficients ${{{\widehat s}_n}}$'s can also be obtained from
\begin{align}\label{eq:HG58}
{{{\widehat s}_n}} =\int\limits_{ - \infty }^\infty  {\int\limits_{ - \infty }^\infty  {S_{\rm NB}(x,y)LG^*_{0,n}(x,y)} }dxdy.
\end{align}
With these coefficients, the input signal can be recovered by (\ref{eq:HG52}).


For digital computation, discrete HG functions and discrete LG functions are required.
Besides, they must be orthonormal to each other.
The discrete HG functions are usually generated by the commuting matrices of the DFT \cite{martucci1994symmetric,candan2000discrete,pei2006discrete,santhanam2007discrete,candan2007higher,pei2008generalized}.
Here, we use the algorithm in \cite{pei2008generalized} which has the lowest approximation error.
There's no approach to directly generate the discrete LG functions.
Fortunately, according to \cite{beijersbergen1993astigmatic}, the LG functions can be expressed in terms of the 2D HG functions, i.e.
\begin{align}\label{eq:HG64}
L{G_{m,n}}(x,y) =\! \sum\limits_{k = 0}^{m + n} {{j^{m + n - k}}d_{\frac{{m + n}}{2} - k,\frac{{n - m}}{2}}^{\frac{{m + n}}{2}}\!\left( {\frac{\pi }{2}} \right)} H{G_k}(x)H{G_{m + n - k}}(y),
\end{align}
where $d_{M,M'}^J(\beta)$ is the Wigner-d function.
When $m=0$, we have
\begin{align}\label{eq:HG68}
LG_{0,n}(x,y) = \sum\limits_{k = 0}^n {{j^{n - k}}d_{\frac{n}{2} - k,\frac{n}{2}}^{\frac{n}{2}}\!\left( {\frac{\pi }{2}} \right)} H{G_k}(x)H{G_{n - k}}(y).
\end{align}
According to (\ref{eq:HG68}), the normalized Bargmann transform in (\ref{eq:HG56}) can be alternatively expressed as the following form:
\begin{align}\label{eq:HG72}
{S_{{\rm{NB}}}}(x,y) &= \sum\limits_{n = 0}^\infty  {{{\hat s}_n}} \sum\limits_{k = 0}^n {{j^{n - k}}d_{\frac{n}{2} - k,\frac{n}{2}}^{\frac{n}{2}}\left( {\frac{\pi }{2}} \right)} H{G_k}(x)H{G_{n - k}}(y)\nn\\
&= \sum\limits_{k + l = 0}^\infty  {{{\tilde s}_{k,l}}} H{G_k}(x)H{G_l}(y),
\end{align}
where the new coefficient ${\tilde s}_{k,l}$ is defined as
\begin{align}\label{eq:HG76}
{\tilde s_{k,l}} = {\hat s_{k + l}}{j^l}d_{\frac{{l - k}}{2},\frac{{k + l}}{2}}^{\frac{{k + l}}{2}}\left( {\frac{\pi }{2}} \right).
\end{align}
Therefore, the normalized Bargmann transform can be completely realized by the HG functions.
For the inverse transform, one can obtain ${\tilde s}_{k,l}$  from ${S_{{\rm{NB}}}}(x,y)$ by
\begin{align}\label{eq:HG80}
{{\tilde s}_{k,l}} = \int\limits_{ - \infty }^\infty  {\int\limits_{ - \infty }^\infty  {{S_{{\rm{NB}}}}(x,y)H{G_k}(x)H{G_l}(y)} } dxdy,
\end{align}
and then obtain ${\hat s_{k + l}}$ from the relationship in (\ref{eq:HG76}) to recover the signal $s(t)$ by (\ref{eq:HG52}).
Assume there are $N$ input samples which form  a vector $\bb s$,
and there are $N$ orthonormal discrete HG functions which form an $N\times N$ orthonormal matrix $\bf H$.
Then, the discrete forms of (\ref{eq:HG54})  and (\ref{eq:HG72}) are given by
\begin{align}\label{eq:HG84}
{\bf{\hat s}} = {\bf{H}^T\bf{s}}\quad \textmd{and}\quad {{\bf{S}}_{{\rm{NB}}}} = {\bf{H\tilde S}}{{\bf{H}}^T},
\end{align}
respectively.
The vector ${\bf{\hat s}}$ consists of the $N$ coefficients ${{\widehat s}_n}$'s.
The $N\times N$ matrix ${\bf{\tilde S}}$ has $N(N + 1)/2$  nonzero elements; that is, the $(l+1,k+1)$-th element is given by
\begin{align}\label{eq:HG88}
{\left[ {{\bf{\tilde S}}} \right]_{l + 1,k + 1}} = \left\{ {\begin{array}{*{20}{l}}
{{{\hat s}_{k + l}}{j^l}d_{\frac{{l - k}}{2},\frac{{k + l}}{2}}^{\frac{{k + l}}{2}}\left( {\frac{\pi }{2}} \right),}&{0 \le k + l \le N - 1}\\
{0,}&\textmd{otherwise}
\end{array}} \right..
\end{align}
And obviously, the discrete form of the inverse transform is given by
\begin{align}\label{eq:HG92}
{\bf{\tilde S}} = {{\bf{H}}^T}{{\bf{S}}_{{\rm{NB}}}}{\bf{H}}\quad \textmd{and}\quad {\bf{s}} = {\bf{H\hat s}}.
\end{align}


\begin{figure}[t]
\centering
\includegraphics[width=0.9\columnwidth,clip=true]{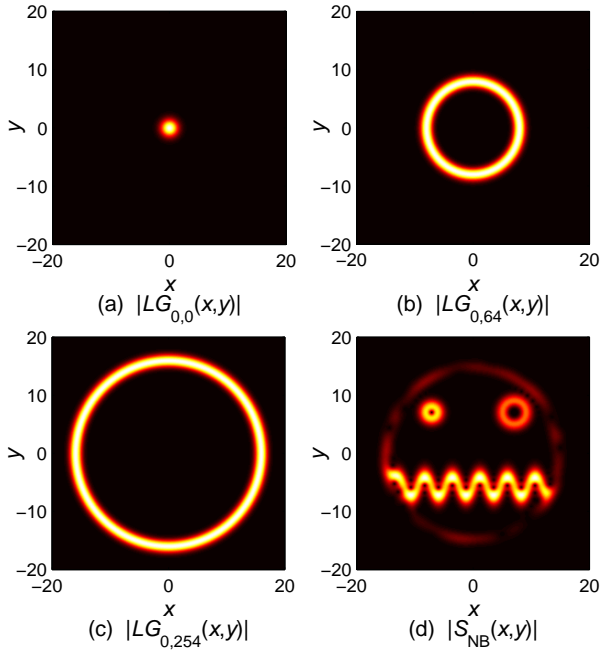}
\vspace*{-4pt}
\caption{
Computation based on HG expansion and synthesis:
(a)-(c) envelops of  
discrete LG functions $LG_{0,0}$, $LG_{0,63}$ and $LG_{0,254}$,
and (d) envelop of the normalized Bargmann transform synthesized by the $255$ discrete LG functions, where $\Delta_x=\Delta_y=\Delta_t=0.157$.
The discrete LG functions are generated by the discrete HG functions according to (\ref{eq:HG68}).
}
\label{fig:HGF}
\end{figure}

An example is given in Fig.~\ref{fig:HGF} where the input signal is the same as that in Fig.~\ref{fig:Gabor}.
The signal has length $N=255$ and sampling period $\Delta_t=0.157$.
Therefore, there are only $255$ discrete HG functions followed by $255$ coefficients ${{\widehat s}_n}$'s, i.e. $n=0,1,\ldots,254$.
Based on (\ref{eq:HG68}), $255$ discrete LG functions can be generated by the discrete HG functions.
Figs.~\ref{fig:HGF} (a)-(c) show the envelopes of
the discrete LG functions $LG_{0,0}$, $LG_{0,63}$ and $LG_{0,254}$, respectively, with $\Delta_x=\Delta_y=0.157$.
The normalized Bargmann transform 
computed by (\ref{eq:HG84}) and (\ref{eq:HG88})
is depicted in Fig.~\ref{fig:HGF} (d).
We can find out the inaccuracy near the boundary because the $255$ discrete LG functions are not enough to approximate the continuous output $S_{\rm NB}(x,y)$.
One approach to avoid this problem is
zero-padding the input signal to increase $N$ and generate more discrete HG and LG functions.
Even though there is some inaccuracy,
the input signal can still be recovered losslessly from the inverse normalized Bargmann transform in 
(\ref{eq:HG92}) and (\ref{eq:HG88}).
The NMSE is below $10^{-25}$ in this example.

\vspace*{18pt}
\section{Computation Based on Gyrator Transform}\label{sec:Gyrator}
The gyrator transform \cite{rodrigo2007gyrator,rodrigo2007experimental}
is a 2D transform
developed to produce rotations in the twisted space-frequency planes.
Given a 2D signal $s(t,\tau)$, the gyrator transform with rotation
angle $\alpha$, denoted by ${\cal G}_\alpha$, is defined as
\begin{align}\label{eq:Gyrator04}
{{\cal G}_\alpha }\left\{ {s(t,\tau )} \right\}  =
\frac{{\left| {\csc \alpha } \right|}}{{2\pi }}\int\limits_{ - \infty }^\infty  \int\limits_{ - \infty }^\infty  {{e^{\frac{{j\left( {xy + t\tau } \right)}}{{\tan \alpha }} - \frac{{j(x\tau  + yt)}}{{\sin \alpha }}}}}  s(t,\tau )dtd\tau .
\end{align}
It has been shown in \cite{rodrigo2007gyrator} that the gyrator transform of the 2D HG function with $\alpha=\pm\frac{\pi}{4}$ is the LG function:
\begin{align}\label{eq:Gyrator08}
{{\cal G}_{\frac{\pi}{4}}}\left\{ HG_n(t)HG_m(\tau) \right\} &= {( - j)^m}L{G_{n,m}}(x,y),\\
{{\cal G}_{ - \frac{\pi}{4}}}\left\{ HG_n(t)HG_m(\tau) \right\} &= {j^m}L{G_{m,n}}(x,y),\label{eq:Gyrator12}
\end{align}
where the HG and LG functions are defined in (\ref{eq:HG04}) and (\ref{eq:HG36}).
Let $\alpha=-\frac{\pi}{4}$ and $m=0$, and then one has
\begin{align}\label{eq:Gyrator16}
\pi ^{ - \frac{1}{4}}{{\cal G}_{ - \frac{\pi}{4}}}\left\{ HG_n(t)e^{ - \frac{\tau ^2}{2}} \right\}= L{G_{0,n}}(x,y).
\end{align}

\begin{figure}[t]
\centering
\includegraphics[width=0.9\columnwidth,clip=true]{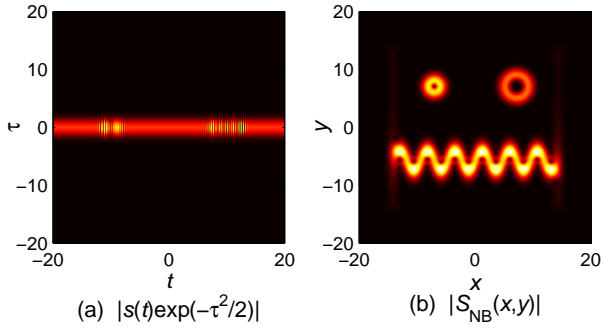}
\vspace*{-4pt}
\caption{
Digital computation based on gyrator transform: (a) envelop of the 2D signal $s(t) e^{ - \frac{\tau ^2}{2}}$ and (b) envelop of the normalized Bargmann transform calculated by the discrete gyrator transform DGT-CCC in \cite{pei2015discrete}.
The sampling periods are $\Delta_\tau=\Delta_x=\Delta_y=\Delta_t=0.157$.
}
\label{fig:DGTCCC}
\end{figure}

According to (\ref{eq:HG44}), the normalized Bargmann transform of the HG function is also the LG function.
Therefore, one has
\begin{align}\label{eq:Gyrator20}
{\cal N\!B}\!\left\{H{G_n}(t)\right\}=\pi ^{ - \frac{1}{4}}{{\cal G}_{ - \frac{\pi}{4}}}\left\{ HG_n(t)e^{ - \frac{\tau ^2}{2}} \right\}.
\end{align}
It follows that (\ref{eq:HG56}) can be rewritten as
\begin{align}\label{eq:Gyrator26}
S_{\rm NB}(x,y)
&= \sum\limits_{n = 0}^\infty  {{{\widehat s}_n}}\pi ^{ - \frac{1}{4}}{{\cal G}_{ - \frac{\pi}{4}}}\left\{ HG_n(t)e^{ - \frac{\tau ^2}{2}} \right\}\nn\\
&= \pi ^{ - \frac{1}{4}}{{\cal G}_{ - \frac{\pi}{4}}}\left\{\sum\limits_{n = 0}^\infty  {{{\widehat s}_n}} HG_n(t)e^{ - \frac{\tau ^2}{2}}\right\} .
\end{align}
Therefore,  the normalized Bargmann transform can be calculated by the gyrator transform:
\begin{align}\label{eq:Gyrator28}
S_{\rm NB}(x,y)=\pi ^{ - \frac{1}{4}}{{\cal G}_{ - \frac{\pi}{4}}}\left\{s(t) e^{ - \frac{\tau ^2}{2}} \right\}.
\end{align}
That is, firstly covert the input signal into a 2D signal by using the  Gaussian function,
and then performing the gyrator transform to the 2D signal with angle $\alpha=-\frac{\pi}{4}$.
For digital computation, a discrete gyrator transform is required.
Several kinds of discrete gyrator transforms have been proposed in \cite{pei2009properties,liu2011fast,pei2015discrete}.
Here, we use the one based on circular chirp convolution, called DGT-CCC, in \cite{pei2015discrete}.
Fig.~\ref{fig:DGTCCC} shows the digital computation of the normalized Bargmann transform based on the gyrator transform.
The input signal in Fig.~\ref{fig:Gabor}(a) is used again, and the sampling period is $\Delta_t=0.157$.
First, convert the input signal $s(t)$ into a 2D signal $s(t) e^{ - \frac{\tau ^2}{2}}$, the envelope of which is depicted in Fig.~\ref{fig:DGTCCC}(a).
Fig.~\ref{fig:DGTCCC}(b) shows the envelope of the normalized Bargmann transform, which is calculated by the DGT-CCC with $\Delta_x=\Delta_y=0.157$.

The gyrator transform has a very simple recovery property, i.e. the inverse gyrator transform with angle $\alpha$ is equivalent to the forward gyrator transform with angle $-\alpha$,
\begin{align}\label{eq:Gyrator30}
\left[\,{\cal G}_{\alpha}\,\right]^{-1}={\cal G}_{-\alpha}.
\end{align}
From (\ref{eq:Gyrator28}) and (\ref{eq:Gyrator30}), one has
\begin{align}\label{eq:Gyrator32}
s(t) e^{ - \frac{\tau ^2}{2}}=\pi ^{ \frac{1}{4}}{{\cal G}_{  \frac{\pi}{4}}}\left\{S_{\rm NB}(x,y) \right\}.
\end{align}
And it follows that $s(t)$ can be easily recovered from
\begin{align}\label{eq:Gyrator40}
s(t) = {\left[ {s(t){e^{ - \frac{{{\tau ^2}}}{2}}}} \right]_{\tau  = 0}} = {\pi ^{\frac{1}{4}}}{\left[ {{G_{\frac{\pi }{4}}}\left\{ {{S_{\rm NB}}(x,y)} \right\}} \right]_{\tau  = 0}}.
\end{align}
In the digital computation of the above inverse transform, the discrete gyrator transform can be simplified because only part of the output (i.e. at $\tau  = 0$) is used.
For the example in Fig.~\ref{fig:DGTCCC}, the input signal recovered from (\ref{eq:Gyrator40}) has NMSE below $10^{-24}$.


\vspace*{18pt}
\section{Computation Based on 2D nonseparable LCT}\label{sec:2DLCT}
The 2D nonseparable LCT (2D NsLCT) \cite{folland1989harmonic,pei2001two,alieva2005alternative}
with $4\times4$ parameter matrix $\bb M=(\bb A,\bb B;\bb C,\bb D)$ is defined as
\begin{align}\label{eq:2DLCT04}
&{\cal O}_{\rm NsLCT}^{\bb M}\{s(\bb t)\}\nn\\
&=\frac{1}{{2\pi \sqrt { - \det ({\bf{B}})} }}\int\limits_{}^{} {{e^{\frac{j}{2}\left( {{{\bf{z}}^T}{\bf{D}}{{\bf{B}}^{ - 1}}{\bf{z}} - 2{{\bf{t}}^T}{{\bf{B}}^{ - 1}}{\bf{z}} + {{\bf{t}}^T}{{\bf{B}}^{ - 1}}{\bf{At}}} \right)}}} s({\bf{t}})d{\bf{t}},
\end{align}
where $\bb t=[t,\tau]^T$ is the input argument, $\bb z=[x,y]^T$ is the output argument, and $s(\bb t)=s(t,\tau)$.
The gyrator transform is a special case of the 2D NsLCT when the parameter matrix is given by
\begin{align}\label{eq:2DLCT08}
\bb M=\bb M_{\alpha}\buildrel \Delta \over= \begin{bmatrix}
{\cos \alpha }&0&0&{\sin \alpha }\\
0&{\cos \alpha }&{\sin \alpha }&0\\
0&{ - \sin \alpha }&{\cos \alpha }&0\\
{ - \sin \alpha }&0&0&{\cos \alpha }
\end{bmatrix}.
\end{align}
According to (\ref{eq:Gyrator28}), the gyrator transform with $\alpha=-\frac{\pi}{4}$ is used in the normalized Bargmann transform.
When $\alpha=-\frac{\pi}{4}$, the above parameter matrix becomes
\begin{align}\label{eq:2DLCT08}
\bb M_{-\pi/4}= \begin{bmatrix}
{\frac{1}{{\sqrt 2 }}}&0&0&{ - \frac{1}{{\sqrt 2 }}}\\
0&{\frac{1}{{\sqrt 2 }}}&{ - \frac{1}{{\sqrt 2 }}}&0\\
0&{\frac{1}{{\sqrt 2 }}}&{\frac{1}{{\sqrt 2 }}}&0\\
{\frac{1}{{\sqrt 2 }}}&0&0&{\frac{1}{{\sqrt 2 }}}
\end{bmatrix},
\end{align}
and the gyrator transform in (\ref{eq:Gyrator28}) can be replaced by the 2D NsLCT with parameter matrix $\bb M_{-\pi/4}$, i.e.
\begin{align}\label{eq:2DLCT12}
S_{\rm NB}(x,y)=\pi ^{ - \frac{1}{4}}{\cal O}_{\rm NsLCT}^{\bb M_{-\pi/4}}\left\{s(t) e^{ - \frac{\tau ^2}{2}} \right\}.
\end{align}

\begin{figure}[t]
\centering
\includegraphics[width=0.9\columnwidth,clip=true]{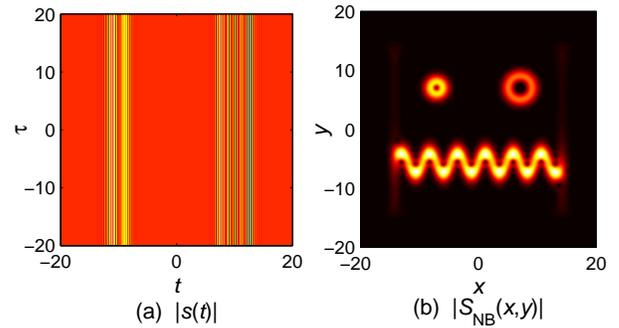}
\vspace*{-4pt}
\caption{
Digital computation based on 2D NsLCT : (a) envelop of the 2D signal $s(t,\tau)=s(t)$ and (b) envelop of the normalized Bargmann transform calculated by the 2D NsDLCT based on CM-CC-CM decomposition.
Sampling periods are $\Delta_t=\Delta_\tau=\Delta_x=\Delta_y=0.157$.
}
\label{fig:2DDLCT}
\end{figure}

When $\bb B=\bb 0$, the definition in (\ref{eq:2DLCT04}) is invalid.
Instead, the 2D NsLCT is defined as
\begin{align}\label{eq:2DLCT16}
{\cal O}_{\rm NsLCT}^{\bb M}\{s(\bb t)\}=\sqrt {\det ({\bf{D}})} {e^{\frac{j}{2}{{\bf{t}}^T}{\bf{C}}{{\bf{D}}^T}{\bf{t}}}}s\!\left( {{{\bf{D}}^T}{\bf{t}}} \right),
\end{align}
where $\bb t=[t,\tau]^T$ and $s(\bb t)=s(t,\tau)$.
The parameter matrix ${\bb M}$ in the 2D NsLCT is real.
If we let ${\bb M}$  be complex as follows
\begin{align}\label{eq:2DLCT20}
\bb M=\bb M_{\rm G} \buildrel \Delta \over =\begin{bmatrix}
1&0&0&0\\
0&1&0&0\\
0&0&1&0\\
0&j&0&1
\end{bmatrix},
\end{align}
the 2D NsLCT in (\ref{eq:2DLCT16}) becomes
\begin{align}\label{eq:2DLCT24}
{\cal O}_{\rm NsLCT}^{\bb M_{\rm G}}\{s(t,\tau)\} = {e^{ - \frac{1}{2}{\tau ^2}}}s(t,\tau ),
\end{align}
which is a multiplication with Gaussian function.
If $s(t,\tau)=s(t)$, the term $s(t){e^{ - \frac{1}{2}{\tau ^2}}}$ in (\ref{eq:2DLCT12}) can be replaced by the above equation, i.e.
\begin{align}\label{eq:2DLCT28}
S_{\rm NB}(x,y)=\pi ^{ - \frac{1}{4}} {\cal O}_{\rm NsLCT}^{\bb M_{-\pi/4}}\left\{{\cal O}_{\rm NsLCT}^{\bb M_{\rm G}}\left\{ s(t,\tau) \right\}\right\}.
\end{align}
Due to the additivity of the 2D NsLCT,
the two 2D NsLCTs in the above equation can be combined into one 2D NsLCT, i.e.
\begin{align}\label{eq:2DLCT32}
S_{\rm NB}(x,y)=\pi ^{ - \frac{1}{4}}{\cal O}_{\rm NsLCT}^{\bb M_{\rm NB}}\left\{ s(t,\tau) \right\},
\end{align}
where $s(t,\tau)=s(t)$ and
$\bb M_{\rm NB}$ is
defined as
\begin{align}\label{eq:2DLCT36}
\bb M_{\rm NB}=\bb M_{-\pi/4}\cdot\bb M_{\rm G}
=\begin{bmatrix}
{\frac{1}{{\sqrt 2 }}}&{ - j\frac{1}{{\sqrt 2 }}}&0&{ - \frac{1}{{\sqrt 2 }}}\\
0&{\frac{1}{{\sqrt 2 }}}&{ - \frac{1}{{\sqrt 2 }}}&0\\
0&{\frac{1}{{\sqrt 2 }}}&{\frac{1}{{\sqrt 2 }}}&0\\
{\frac{1}{{\sqrt 2 }}}&{j\frac{1}{{\sqrt 2 }}}&0&{\frac{1}{{\sqrt 2 }}}
\end{bmatrix}.
\end{align}
Several kinds of 2D nonseparable discrete LCT (NsDLCT) have been proposed in \cite{kocc2010fast,ding2012improved,pei2016two}.
In Fig.~\ref{fig:2DDLCT}, we use the 2D NsDLCT based on CM-CC-CM decomposition to compute the normalized Bargmann transform.
First, the input signal $s(t)$ is treated as a 2D signal, as shown in Fig.~\ref{fig:2DDLCT}(a).
Next, perform the 2D NsDLCT with parameter matrix $\bb M_{\rm NB}$ given in (\ref{eq:2DLCT36}), and then we have the normalized Bargmann transform with envelope depicted in Fig.~\ref{fig:2DDLCT}(b).
Sampling periods $\Delta_t=\Delta_\tau=\Delta_x=\Delta_y=0.157$ are used.

Like the gyrator transform, the 2D NsLCT also has the reversibility property that the inverse of the 2D NsLCT with parameter matrix $\bb M$ is equivalent to the 2D NsLCT with parameter matrix $\bb M^{-1}$:
\begin{align}\label{eq:Gyrator30}
\left[\,{\cal O}_{\rm NsLCT}^{\bb M}\,\right]^{-1}={\cal O}_{\rm NsLCT}^{\bb M^{-1}}.
\end{align}
Therefore, the inverse normalized Bargmann transform can also be realized by the 2D NsLCT:
\begin{align}\label{eq:2DLCT40}
s(t)  = {\pi ^{\frac{1}{4}}}{\left[
{\cal O}_{\rm NsLCT}^{\bb M_{\rm NB}^{-1}}\left\{ {S_{\rm NB}}(x,y) \right\}
\right]_{\tau  = 0}}.
\end{align}
For digital computation, the used 2D NsDLCT can be simplified because we only require the output data at $\tau  = 0$.
We recover the input signal for the example in Fig.~\ref{fig:2DDLCT} by the 2D NsDLCT based on CM-CC-CM decomposition, and the NMSE is  below $10^{-31}$.

\begin{figure}[t]
\centering
\includegraphics[width=0.98\columnwidth,clip=true]{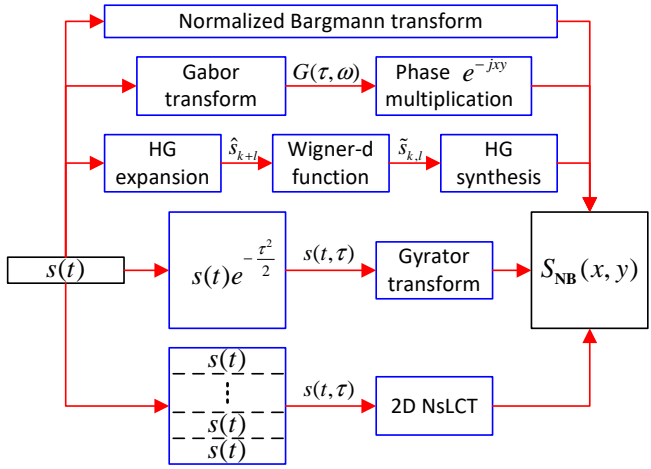}
\vspace*{-4pt}
\caption{
Block diagrams of the four proposed methods.
}
\label{fig:Block}
\end{figure}

\vspace*{18pt}
\section{Computational Complexity and Accuracy}\label{sec:ComAcc}
Before analyzing complexity and accuracy of the four proposed methods, we briefly summarize their concepts in Fig.~\ref{fig:Block}.

\subsection{Computational Complexity}\label{subsec:Com}
Assume there are $N$ input samples with complex values,
and for simplicity, assume the size of the discrete output is $N\times N$.
We also assume that all the kernel functions, window functions and matrices are computed in advance and stored in memory if they are precomputable.
In this section, we use the number of real multiplications in each method to evaluate the computational complexity.

A straightforward method to implement the normalized Bargmann transform is sampling it into the following discrete form:
\begin{align}\label{eq:Com04}
&{S_{{\rm{NB}}}}(p{\Delta _x},q{\Delta _y})\nn\\
&= {\pi ^{ - \frac{3}{4}}}{\Delta _t}\sum\limits_n^{} {e^{ - {p^2}\Delta _x^2 - jpq{\Delta _x}{\Delta _y} + \sqrt 2 (p{\Delta _x} + jq{\Delta _y})n{\Delta _t} - \frac{{{n^2}\Delta _t^2}}{2}}} s(n{\Delta _t}),
\end{align}
and directly calculating the summation.
We call it direct summation method.
Each output sample requires $N$ complex multiplications.
Accordingly, the direct summation method requires  $N^3$ complex multiplications, i.e. $4N^3$ real multiplications.

Recall the Gabor-based method in (\ref{eq:Gabor16}).
We need $N\times N$ discrete output of the Gabor transform, i.e. $N$ time samples and $N$ frequency samples.
For each time sample,  one Gaussian window multiplication and one FFT are used if the FFT-based algorithm is used to implement the Gabor transform.
Thus, there are $2N$ real multiplications (because the Gaussian window is real) and  $\frac{{{N}}}{2}{\log _2}N$ complex multiplications at each time sample.
At last, another $N^2$ complex multiplications are used to obtain the normalized Bargmann transform from the Gabor transform.
Thus, the total number of real multiplications required by the Gabor-based method is
\begin{align}\label{eq:Com08}
N \cdot \left( {2N + 4 \cdot \frac{N}{2}{{\log }_2}N} \right) + 4 \cdot {N^2} = 2{N^2}{\log _2}N + 6{N^2}.
\end{align}

Next, consider the discrete form of the HG-based method in (\ref{eq:HG84}) and (\ref{eq:HG88}).
Because $\bb H$ is real, one only needs $2N^2$ real multiplications to compute $\bf{\hat s}={\bf{H}^T\bf{s}}$.
Given $\bf{\hat s}$, one can obtain the $N(N + 1)/2$  nonzero elements in ${{\bf{\tilde S}}}$ (i.e. $0\leq k+l\leq N-1$) by $N(N + 1)/2$ complex multiplications.
And because $\bb H$ is real and there are many zeros in ${{\bf{\tilde S}}}$, the matrix multiplication of $\bf{H}$ and $\bf{\tilde S}$ requires only $N^2(N + 1)$ real multiplications,
while the matrix multiplication of ${\bf{H\tilde S}}$ and ${\bf{H}}^T$ needs $2N^3$ real multiplications.
Thus, we can conclude that the HG-based method requires
\begin{align}\label{eq:Com12}
2{N^2} + 4 \cdot \frac{{N(N + 1)}}{2} + {N^2}(N + 1) + 2{N^3} = 3{N^3} + 5{N^2} + 2N
\end{align}
real multiplications.

For the gyrator-based method in (\ref{eq:Gyrator28}),
one needs $2N^2$ real multiplications to convert the 1D input signal $s(t)$ into the 2D signal $s(t) e^{ - \frac{\tau ^2}{2}}$.
Therefore, the complexity of the gyrator-based method is equal to the complexity of the gyrator transform plus $2N^2$ real multiplications.
As in Fig.~\ref{fig:DGTCCC}, if the DGT-CCC in \cite{pei2015discrete} is employed to compute the gyrator transform, the gyrator-based method totally requires
\begin{align}\label{eq:Com16}
2{N^2} + 12{N^2} + 4{N^2}{\log _2}{N^2} = 8{N^2}{\log _2}N + 14{N^2}
\end{align}
real multiplications.

Consider the 2D NsLCT-based method in (\ref{eq:2DLCT32}).
First, one needs to clone the 1D input signal into a 2D signal, i.e. $s(t,\tau)=s(t)$, the complexity of which is negligible.
Therefore, the complexity of the 2D NsLCT-based method is almost equal to the complexity of the 2D NsLCT.
If we use the 2D NsDLCT based on CM-CC-CM decomposition as in Fig.~\ref{fig:2DDLCT}, the number of real multiplications is
\begin{align}\label{eq:Com20}
4 \cdot \left( {{N^2}{{\log }_2}{N^2} + 3{N^2}} \right) = 8{N^2}{\log _2}N + 12{N^2}.
\end{align}
Comparing (\ref{eq:Com16}) and (\ref{eq:Com20}), we can find out that the gyrator-based method has a little higher complexity because  additional $2N^2$ real multiplications are used when converting the 1D signal into 2D.

We summarize the complexity of the direct summation method and the four proposed methods in Table~\ref{tab:Com}.
Note that any existing fast algorithms of the Gabor transform, gyrator transform and 2D NsLCT can be utilized instead.
But generally, the Gabor-based method has the lowest complexity because the Gabor transform is an 1D-to-1D transform while the gyrator transform and the 2D NsLCT are 2D-to-2D transforms.
The 2D NsLCT-based method has a little lower complexity than the gyrator-based method if similar algorithms are adopted.
For example, in this paper, the algorithms based on CM-CC-CM decomposition are adopted in these two methods.

\begin{table}[t]
\small
\begin{center}
\setstretch{1.5}
\caption{Complexity of the direct summation method and the four proposed methods}\label{tab:Com}
\begin{tabular}{|l|l|}
\hline
 &  Complexity (number of real multiplications)\\
\hline\hline
Direct summation\!\!\! & $4N^3$ \\
\hline
Gabor-based  & $2{N^2}{\log _2}N + 6{N^2}$\ \ \ (i.e. Gabor transform$+4{N^2}$)\\
\hline
HG-based  & $3{N^3} + 5{N^2} + 2N$  \\
\hline
gyrator-based  & $8{N^2}{\log _2}N + 14{N^2}$\ (i.e. gyrator transform$+2N^2$)\!\!\!\\
\hline
2D NsLCT-based  & $8{N^2}{\log _2}N + 12{N^2}$\ (i.e. 2D NsLCT)\\
\hline
\end{tabular}
\end{center}
\end{table}

\subsection{Accuracy}\label{subsec:Acc}
\begin{figure}[t]
\centering
\includegraphics[width=0.9\columnwidth,clip=true]{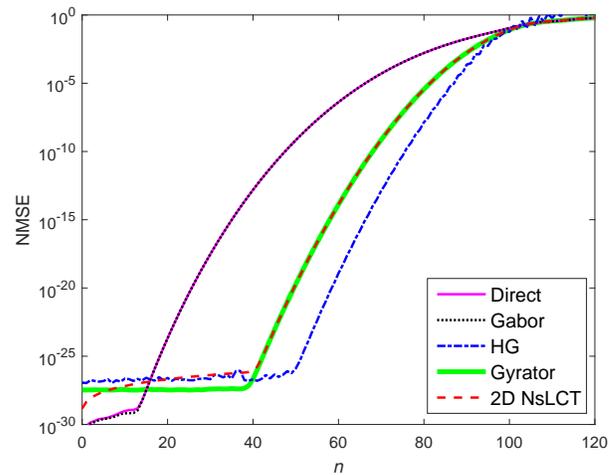}
\vspace*{-4pt}
\caption{
Accuracy of the direct summation method and the four proposed computational methods.
The $127$-point sampled HG functions with order $n=0,1,\ldots,120$ are used as the input, and the sampling periods are $\Delta_t=\Delta_x=\Delta_y=0.2224$.
}
\label{fig:Acc}
\end{figure}

It has been mentioned in (\ref{eq:HG44}) that the normalized Bargmann transform of the HG function is the LG function, i.e.
\begin{align}\label{eq:Acc04}
{\cal N\!B}\!\left\{H{G_n}(t)\right\}=L{G_{0,n}}(x,y).
\end{align}
With sampled HG function as the discrete input, we want the discrete output of the digital computation can approximate the sampled LG function.
A more accurate computational method should have smaller approximation error.
Sampling the HG and LG function with $\Delta_t=\Delta_x=\Delta_y=0.2224$ and length $N=127$, the approximation errors of the
direct summation method and the four proposed computational methods  are shown in Fig.~\ref{fig:Acc}.
The order $n$ of the HG function ranges from 0 to 120.
For all the methods, the accuracy drops when $n$ increases.
The Gabor-based method has the lowest accuracy in most cases, but it also has the lowest computational complexity.
The direct summation method has almost the same accuracy as the Gabor-based method.
The gyrator-based and the 2D NsLCT-based methods have similar accuracy because both of them are computed by algorithms based on the CM-CC-CM decomposition in this paper.
The HG-based method has the highest accuracy
when $n\leq100$, but on the contrary has the lowest accuracy when $n>100$.
Next, we will discuss how to determine $N$, $\Delta_t$, $\Delta_x$ and $\Delta_y$ to achieve high accuracy.

The normalized Bargmann transform is just the Gabor transform multiplied by some phase term.
If the sampling rate is lower than the Nyquist rate, aliasing occurs, i.e. overlapping in frequency domain.
It follows that the discrete output of the normalized Bargmann transform will have overlapping effect in the $y$ axis, no matter what kind of method is utilized.
In Fig.~\ref{fig:Acc}, when $n$ increases, the accuracy decreases because the input (i.e. the HG function) has larger bandwidth and leads to greater inaccuracy on the boundary of $y$ axis.
The HG functions are approximately bandlimited, and thus high accuracy can be achieved if the sampling interval is small enough.
However, it is possible that the input signal is not bandlimited or approximately bandlimited, requiring sampling interval infinitely small, such as chirp signals.
To solve this problem, one can try separating the signal into smaller segments such that each segment would be approximately bandlimited.

Another reason why the accuracy decreases when $n$ increases is the time duration of the input signal.
If the input signal is not time-limited to $N\Delta_t$,   inaccuracy on the boundary of $x$ axis is inevitable.
Nevertheless, from Figs.~\ref{fig:Gabor}-\ref{fig:2DDLCT}, we can find out that the proposed methods have good enough performance except the boundary part.
Even if $\Delta_t$ is small enough and $N$ is large enough, the HG-based method has another problem, as shown in Fig.~\ref{fig:HGF}.
This is because the limited number of (i.e. $N$) discrete LG functions are not enough to well represent the sampled output.
And this is why in Fig.~\ref{fig:Acc}, the HG-based method is worse than the other proposed method when $n>100$.
Thus, one may needs to zero-pad the input, i.e. further increase $N$,
to increase the number of discrete LG functions.

Generally, the output sampling intervals $\Delta_x$ and $\Delta_y$
won't affect the accuracy, but
there may be some restrictions depending on what algorithm one is using.
For example, in the gyrator-based method, one needs ${\Delta _x} = {\Delta _t}$ and ${\Delta _y} = {\Delta _\tau }$ if the DGT-CCC in \cite{pei2015discrete} is employed.
If another algorithm, DGT-LCC in \cite{pei2015discrete}, is used instead, $\Delta_x$ and $\Delta_y$ can be
arbitrary, but the cost is higher computational complexity.

\section{Conclusion and Future Work}\label{sec:Con}
The Bargmann transform is a special case of the complex LCT.
Because the output of the Bargmann transform may be unbounded near infinity, the normalized Bargmann transform is considered.
We derive the relationships of the normalized Bargmann transform to the Gabor transform,  the Hermite Gaussian functions, the gyrator transform and the 2D nonseparable LCT.
Four kinds of computational methods of the normalized Bargmann transform are proposed based on these relationships.
We also derive several computational methods for the inverse normalized Bargmann transform.
If the input signal is time-limited and approximately bandlimited, these computational methods have very high accuracy.

In this paper, we have shown that the normalized Bargmann transform is related to the special case of the 2D NsLCT.
Thus, there may be some connection between the normalized complex LCT and the 2D NsLCT, which may probably be developed in our future work.

\section*{Funding}
Ministry of Science and Technology, Taiwan
(MOST) (MOST 104-2221-E-002-096-MY3, MOST 104-
2221-E-002-006, MOST 104-2917-I-002-042).


\end{document}